\documentclass{article}
\usepackage{LaThuileFPSpro}
\begin{document}
\title{
  Results on axion physics from the CAST experiment at CERN
  }
\author{
  Christos Eleftheriadis        \\
  {\em Nuclear Physics and Elementary Particles Section} \\
  {\em Aristotle University of Thessaloniki}\\
  {\em for the CAST Collaboration$~^*$}}
\maketitle

\baselineskip=11.0pt

\begin{abstract}
Axions are expected to be produced in the sun via the Primakoff
process. They may be detected through the inverse process in the
laboratory, under the influence of a strong magnetic field, giving
rise to X--rays of energies in the range of a few keV. Such an
Axion detector is the CERN Axion Solar Telescope (CAST),
collecting data since 2003. Results have been published, pushing
the axion-photon coupling  g$_{a\gamma}$ below the 10$^{-10}$
GeV$^{-1}$ limit at 95\% CL, for axion masses less than 0.02 eV.
This limit is nearly an order of magnitude lower than previous
experimental limits and surpassed for the first time limits set
from astrophysical arguments based on the energy-loss concept. The
experiment is currently exploring axion masses in the range of
0.02 eV $< m_a <$ 1.1 eV. In the next run, currently under
preparation, the axion mass explored will be extended up to the
limit of 1.1 eV, testing for the first time the region of
theoretical axion models with the axion helioscope method.
\end{abstract}
\long\def\symbolfootnote[#1]#2{\begingroup%
\def\thefootnote{\fnsymbol{footnote}}\footnote[#1]{#2}\endgroup}

\symbolfootnote[0]{$^*$ S.~Andriamonje$^{2}$, S.~Aune$^{2}$,
K.~Barth$^{1}$, A.~Belov$^{12}$, B.~Beltr\'an$^{7}$,
H.~Br\"auninger$^{6}$, J.~Carmona$^{7}$, S.~Cebri\'an$^{7}$,
J.~I.~Collar$^{8}$, T.~Dafni$^{4,2}$, M.~Davenport$^{1}$,
L.~Di~Lella$^{1}$, C.~Eleftheriadis$^{9}$, J.~Englhauser$^{6}$,
G.~Fanourakis$^{10}$, E.~Ferrer-Ribas$^{2}$, H.~Fischer$^{11}$,
J.~Franz$^{11}$, P.~Friedrich$^{6}$, T.~Geralis$^{10}$,
I.~Giomataris$^{2}$, S.~Gninenko$^{12}$, H.~G\'omez$^{7}$,
M.~Hasinoff$^{13}$, F.~H.~Heinsius$^{11}$,
D.~H.~H.~Hoffmann$^{4,5}$, I.~G.~Irastorza$^{2,7}$,
J.~Jacoby$^{14}$, K.~Jakov\v{c}i\'{c}$^{16}$, D.~Kang$^{11}$,
K.~K\"onigsmann$^{11}$, R.~Kotthaus$^{15}$, M.~Kr\v{c}mar$^{16}$,
K.~Kousouris$^{10}$, M.~Kuster$^{4,6}$, B.~Laki\'{c}$^{16}$,
C.~Lasseur$^{1}$, A.~Liolios$^{9}$, A.~Ljubi\v{c}i\'{c}$^{16}$,
G.~Lutz$^{15}$, G.~Luz\'on$^{7}$, D.~Miller$^{8}$,
A.~Morales$^{7}$, J.~Morales$^{7}$, A.~Nordt$^{4,6}$,
A.~Ortiz$^{7}$, T.~Papaevangelou$^{1}$, A.~Placci$^{1}$,
G.~Raffelt$^{15}$, H.~Riege$^{1}$, A.~Rodr\'iguez$^{7}$,
J.~Ruz$^{7}$, I.~Savvidis$^{9}$, Y.~Semertzidis$^{17}$,
P.~Serpico$^{15}$, L.~Stewart$^{1}$, J.~Villar$^{7}$,
J.~Vogel$^{11}$, L.~Walckiers$^{1}$, K.~Zioutas$^{17,1}$
~\\
~\\
1. European Organization for Nuclear Research (CERN), Gen\`eve, Switzerland\\
2. DAPNIA, Centre d'\'Etudes Nucl\'eaires de Saclay (CEA-Saclay), Gif-sur-Yvette, France\\
3. Department of Physics, Queen's University, Kingston, Ontario\\
4. Technische Universit\"{a}t Darmstadt, IKP, Darmstadt, Germany\\
5. Gesellschaft f\"ur Schwerionenforschung, GSI Darmstadt,
Germany\\
6. Max-Planck-Institut f\"{u}r extraterrestrische Physik, Garching, Germany\\
7. Instituto de F\'{\i}sica Nuclear y Altas Energ\'{\i}as, Universidad de Zaragoza, Zaragoza, Spain\\
8. Enrico Fermi Institute and KICP, University of Chicago, Chicago, IL, USA\\
9. Aristotle University of Thessaloniki, Thessaloniki, Greece\\
10. National Center for Scientific Research ``Demokritos'', Athens, Greece\\
11. Albert-Ludwigs-Universit\"{a}t Freiburg, Freiburg, Germany\\
12. Institute for Nuclear Research (INR), Russian Academy of Sciences, Moscow, Russia\\
13. Department of Physics and Astronomy, University of British Columbia, Department of  Physics, Vancouver, Canada\\
14. Johann Wolfgang Goethe-Universit\"at, Institut f\"ur Angewandte Physik, Frankfurt am Main, Germany\\
15. Max-Planck-Institut f\"{u}r Physik (Werner-Heisenberg-Institut), Munich, Germany\\
16. Rudjer Bo\v{s}kovi\'{c} Institute, Zagreb, Croatia\\
17. Physics Department, University of Patras, Patras, Greece    \\
}

\newpage
\section{Introduction}
CP violating terms in quantum chromodynamics give rise to a
non-vanishing neutron electric dipole moment, EDM. However,
experimental efforts have put tight upper limits, $d_n$ $<$ 2.9
$\cdot$10$^{-26}$ e $\cdot$ cm \cite{RAL-EDM}, which is orders of
magnitude more strict than the prediction of theory. The question
of CP conservation in QCD, which is known as the strong CP problem
(SCPP), can be answered through the existence of at least one
massless quark, a hypothesis which is experimentally excluded,
since all quarks have mass \cite{PDG}. Up to now, the most
convincing solution to the SCPP was given by Peccei and Quinn
\cite{PQ1}, \cite{PQ2} through the introduction of a new global
U(1) symmetry, which is spontaneously broken at an energy scale
$f_a$. Through this process, the CP violation in strong
interactions is dynamically suppressed. According to the
Nambu-Goldstone theorem, the break down of the symmetry generates
a Nambu-Goldstone boson, a spinless particle, the axion. Axions
are expected to be much alike pions. If they exist, they should
interact very weakly, being also very light particles. Depending
on their density and mass, they may constitute a candidate for the
cold dark matter in the universe. Axion parameters, namely their
mass and PQ symmetry breaking scale are related through the
following expression

\begin{equation}
m_a= 6\,\, eV \,\, \frac{10^6 \, GeV}{f_a}
 \label{eq:eq1}
\end{equation}
where $f_a$  is the axion decay constant or breaking scale of the
Peccei--Quinn symmetry and $m_a$ is the axion mass.

Axions are also expected to be copiously produced in stellar cores
through their coupling to plasma photons, with energies in the
range of keV. Since their coupling is small, they escape nearly
freely, carrying away amounts of energy from the star. This
dissipation mechanism, if present, increases the rate at which the
stars consume their fuel, in order to counterbalance the axion
energy loss. For supernovae environments the axion energy may
reach even 160 MeV. They constitute again an energy dissipation
mechanism. In general, in all stellar objects, from white dwarfs
to horizontal branch stars, energy dissipation by axions add a new
energy loss channel which can affect the evolution timescale of
these objects and, therefore, their apparent number density on the
sky \cite{raffelt}.

Searches for axions are intense nowadays, including not only the
idea of the helioscope \cite{Sikivie1, Lazarus} presently used by
CAST \cite{CAST-proposal, xrh-gravity, CAST-first-results}, but
also Bragg scattering \cite{Bragg-Paschos, solax, cosme}, cavity
searches \cite{cavity}, the PVLAS experiment method \cite{pvlas,
cameron, raffelt-pvlas}, the "through the wall" \cite{cameron,
ringwald, sikivie2}or even "through the sun" method \cite{quasar}.
Astrophysical and cosmological arguments are involved in order to
shed light on their characteristic parameters \cite{raffelt,
sikivie3}. Solar mysteries may also be explained in terms of
axions \cite{zioutas-sun1}. Axions with earth origin have been
also discussed \cite{tasos}.

CAST is designed to measure axions, produced by the Primakoff
effect in the stellar plasma of the central area of our sun. Other
potential sources of axions may also become of interest in the
future.

\section{Axion production in the sun}
The dominant mechanism in axion production is the conversion of a
plasma photon into an axion, in the field of a charged particle.
Other contributions, such as the "electro-Primakoff" effect, are
not important, because all charged particles in the sun are not
relativistic and, therefore, are not able to provide high B
fields. Photons of energy E in a stellar plasma may be transformed
into axions through the Primakoff effect at a rate given by

\begin{equation}
  \Gamma_{\gamma \rightarrow \alpha} = \frac{g_{\alpha \gamma}^2 T \kappa_s^2 }{32
  \pi} \, \biggl[\biggl(1+\frac{\kappa^2_s}{4E^2} \biggr)\, ln \biggl(1+ \frac{4E^2}{\kappa_s^2} \biggr)-1   \biggr]
  \label{eq:rate}
\end{equation}

In this relation, natural units have been used. T is temperature and $\kappa$ is the screening scale in
the Debye-Huckel approximation \cite{jcap}.

\begin{equation}
\kappa_s^2 = \frac{4 \pi \alpha}{T} \biggl(n_e +
\sum_{nuclei}{Z_j^2 n_j} \biggr)
  \label{eq:DH}
\end{equation}

A discussion on the solar axion flux on earth can be found in reference \ref{l:jcap},
where also the dependence of flux on two different solar
models is given.

\section{The CAST experiment}

The experiment uses a recycled superconducting test magnet from
LHC, with a length of 9.26 meters and can reach a magnetic field
of 9 T at 13 kA. The magnet has two pipes, as all LHC magnets,
with a cross sectional area of 14.5 cm$^2$ each. It is mounted on
a moving structure and it may track the sun for nearly 3 hours a
day, half of the time during morning and half in the evening. This
limitation comes from the fact that it is not possible to increase
the elevation angle of the superconducting magnet more than
$\pm$8$^o$. Coverage of azimuthal angle is 100$^o$. On both ends
of the magnet, on all four apertures X--ray detectors are mounted.
Namely, on the front side, looking for X--rays during sunset,
there is a conventional Time Projection Chamber (TPC) detector
\cite{tpc} with an area covering both holes. On the other side,
one aperture is covered by a position sensitive gaseous micromegas
detector (MM)\cite{mm}, whereas on the second aperture an X--ray
telescope \cite{xray} \cite{ccd} is mounted. Solar axions with an
energy spectrum peaking at 4.2 keV, are transformed into photons
via time reversed Primakoff effect, under the influence of the
transverse magnetic field of 9 T. The conversion probability is
given by

\begin{equation}
P_{a \rightarrow \gamma}= \biggl(\frac{g_{a\gamma}B}{q} \biggr)^2
\,\, \sin^2{ \biggl(\frac{qL}{2}\biggr)}
  \label{eq:prob}
\end{equation}

$q$= $m_a^2$/$2 E$ being the momentum difference between axion and
photon. These photons are expected to be recorded by the three
detectors as signal over background, only when the sun and the
magnet are aligned. The rest of the time the detectors are
measuring pure background. All detectors are able to measure
background also simultaneously with the signal, since they are
position sensitive and their effective area is bigger than the
aperture of the magnet. As a matter of fact, whatever is recorded
outside the area of the detector covering the magnet's aperture,
is pure background even at the time of the alignment of the magnet
to the sun. This is especially true for CCD, since the area of the
focal point of the x-ray telescope is very small, less than 10
mm$^2$. It is an evident advantage to measure background and
signal at the same time, leading to reduced systematics. A
detailed description of the experiment, as well as of the
detectors and the x-ray telescope can be found in references
\ref{l:jcap}, \ref{l:tpc}, \ref{l:mm}, \ref{l:xray}, \ref{l:ccd},
\ref{l:xray-sim-auth}, \ref{l:ccd-cebrian}.

\section {Results and discussion}
\begin{figure}[t]
  \vspace{9.0cm}
  \includegraphics{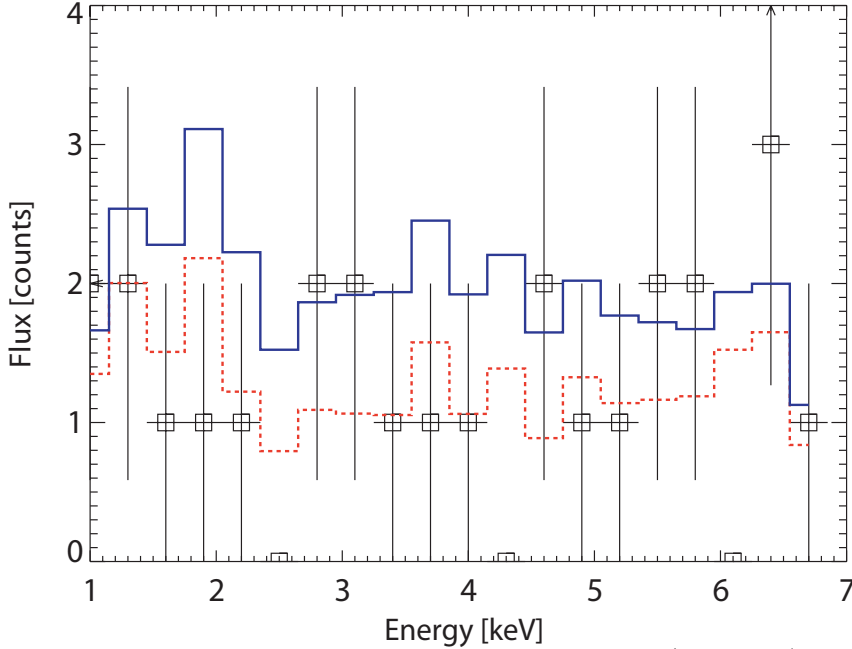}
  \caption{\it
    Spectral distribution of the CCD data (rectangles), expectation for the best
    fit g$_{a \gamma}$ (dashed line) and expectation for the 95\% CL limit
    on g$_{a \gamma}$ (continuous line), in units of counts per energy bin in the spot area (9.35 mm$^2$).
    \label{fig-ccd} }
\end{figure}
All three CAST detectors were collecting data during 2003 and
2004, the sun tracking data being collected for ~300 hours and the
background data for an order of magnitude more time. All three
detectors were significantly improved from 2003 to 2004. In the
system x--ray telescope --- CCD detector, the pointing stability
of the x--ray telescope was continuously monitored and allowed to
reduce the area of the detector where the axion signal was
expected, by a factor of 5.8. This was an essential improvement
for the signal to noise ratio, since the same expected signal was
concentrated in a much smaller area. The integrated background in
the spot area, which is now smaller, is consequently reduced by
the factor of 5.8 mentioned above. Moreover, with better
shielding, the specific background level dropped by another factor
of 1.5. Detailed information on the specifics of the x--ray
telescope and the CCD detector can be found in references
\ref{l:xray},\ref{l:ccd} and \ref{l:xray-sim-auth}. The data set,
shown in fig \ref{fig-ccd} collected during 2004 allowed us to
derive a lower upper limit on the axion--photon coupling. The
analysis procedure is thoroughly described in reference
\ref{l:jcap}. The result for axion-photon coupling g$_{a\gamma}$
is an upper limit of 8.9$\cdot$10$^{-11}$ GeV$^{-1}$, at the 95\%
CL.

\begin{figure}[t]
  \vspace{9.0cm}
  \includegraphics{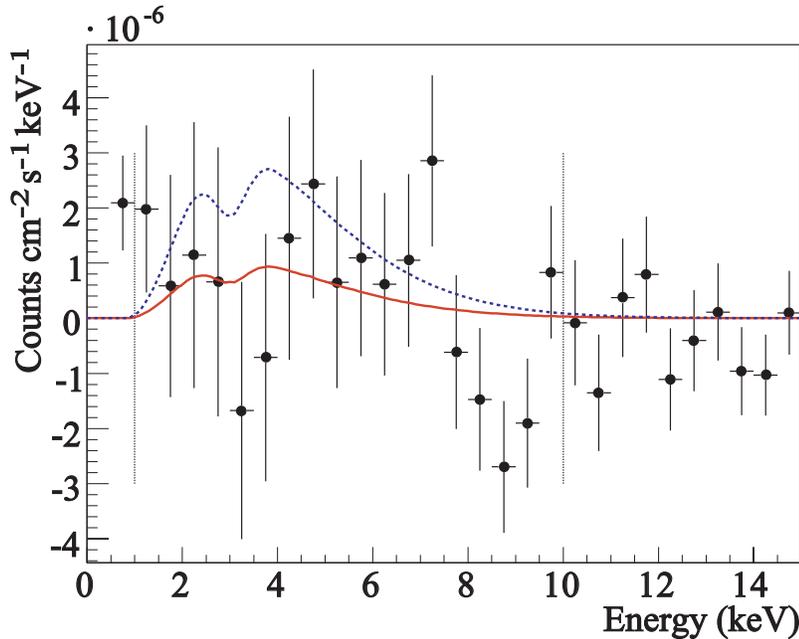}
  \caption{\it
    Experimental subtracted spectrum (bullets), expectation for the best
    fit g$_{a \gamma}$ (continuous line) and expectation for the 95\% CL limit
    on g$_{a \gamma}$ (dashed line), for the TPC data.
    \label{fig-tpc} }
\end{figure}

The TPC detector, looking for sunset axions,  was housed in a new
shielding, consisting of a 5 mm thick copper box which was inside
successive shielding layers of 22 cm of polypropylene, 1 mm of
Cadmium and 2.5 cm of Lead. Care has been taken so that all these
materials were of low radioactivity. Permanent flushing with
nitrogen was creating an overpressure, pushing away any radon
contamination in the area of the detector. The background level
was succeeded to be reduced by a factor of 4.3. Detailed
information on the TPC detector can be found in references
\ref{l:jcap} and \ref{l:tpc}. The results are shown in figure
\ref{fig-tpc}. The upper limit on g$_{a\gamma}$ from the TPC data
for 2004, is 1.29$\cdot$10$^{-10}$ GeV$^{-1}$, at the 95\% CL.

\begin{figure}[t]
  \vspace{9.0cm}
  \includegraphics{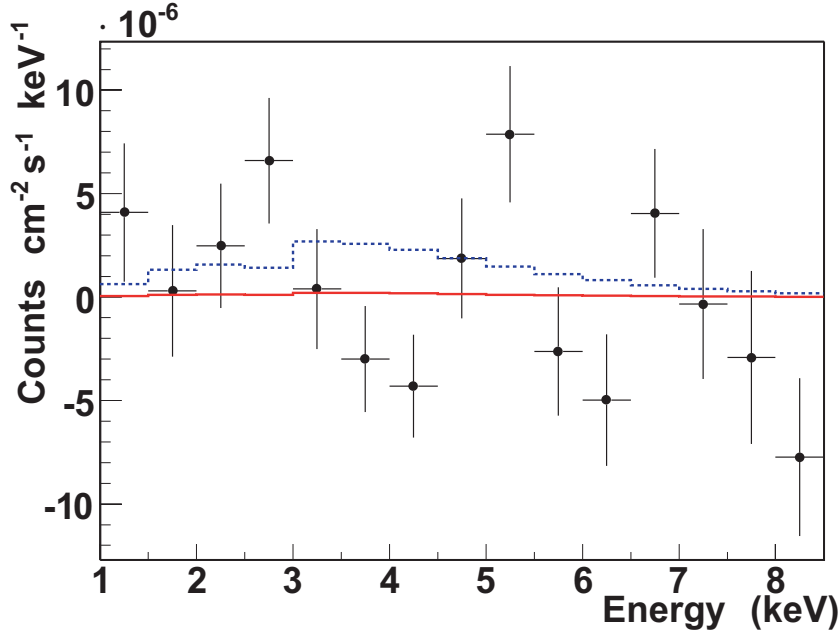}
  \caption{\it
    Experimental subtracted spectrum (bullets), expectation for the best
    fit g$_{a \gamma}$ (continuous line) and expectation for the 95\% CL limit
    on g$_{a \gamma}$ (dashed line), for the Micromegas data.
    \label{fig-mm} }
\end{figure}
The Micromegas detector was placed on the west end of the magnet,
looking for sunrise axions. The newly designed version of the
detector operated smoothly during 2004 data taking and the
analysis technique has been also improved, resulting in a
background level, suppressed by a factor of 2.5 compared to the
2003 data set \cite{mm, jcap}. These improvements allowed to set
an upper limit to  g$_{a\gamma}$ from the MM 2004 data of
1.27$\cdot$10$^{-10}$ GeV$^{-1}$, at the 95\% CL. Taking into
account all three detectors and the data sets of both years 2003
and 2004, for axion masses below 0.02 eV, we obtained a final
upper limit of

\[ g_{a\gamma}  < 8.8 \cdot 10^{-11} GeV^{-1} \,\,\,    (95\%\,\,\, CL)\]

\begin{figure}[t]
  \vspace{9.0cm}
  \includegraphics{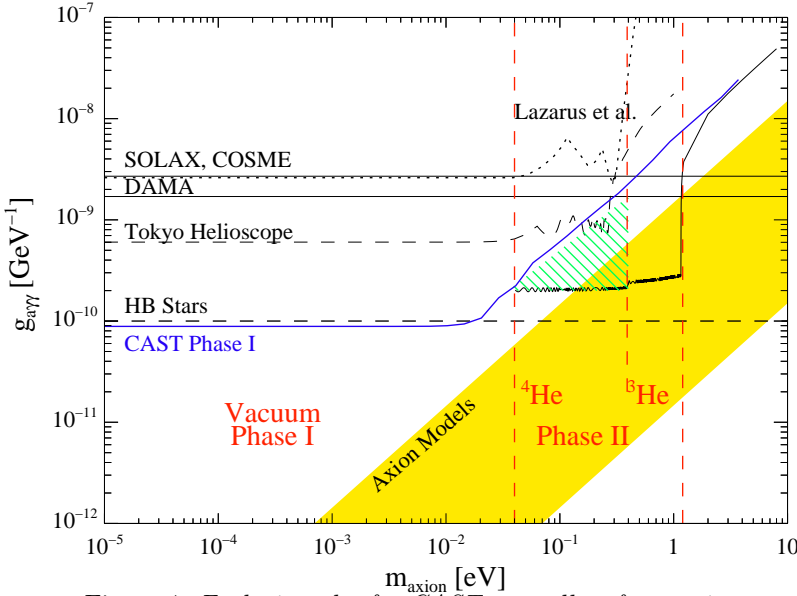}
  \caption{\it
    Exclusion plot for CAST, as well as for previous experiments. Limits from astrophysical
    and cosmological arguments are also shown. The corridor of axion models below 1 eV is checked for the first time.
    \label{fig-ex-plot} }
\end{figure}

For higher axion masses the axion photon-coherence is lost since
their oscillation length is reduced. The exclusion plot in figure
\ref{fig-ex-plot} shows the CAST result together with results from
previous experiments, as well as limits derived from astrophysical
and cosmological arguments. For the first time an experimentally
set limit is better than the one given from arguments based on the
population of Horizontal Branch stars in globular clusters
\cite{jcap}.

\section {Prospects - CAST phase II}

For low axion masses, the axion - photon oscillation length
exceeds by far the length of the magnet, that is axions and
virtual photons are travelling coherently and the axion - photon
transformation probability depends on $B^2L^2$. In this low axion
mass region, recoil effects in the Primakoff effect (and its
inverse effect) may be neglected and the energies of both
particles are considered to be the same. However, at higher axion
masses, the axion - photon coherence is lost due to the axion mass
which prevents it from travelling in phase with virtual photons in
the transverse magnetic field. In order to restore the coherence
condition, we fill the magnet channels with gas, so that the
photon acquires an effective mass $m_{\gamma} > 0$. The momentum
transfer becomes

\begin{equation}
q= \frac{m_a^2 - m_{\gamma}^2}{2E}
  \label{eq:momtra1}
\end{equation}

 as opposed to

\begin{equation}
q= \frac{m_a^2}{2E}
  \label{eq:momtra2}
\end{equation}

The conversion probability in gas is given by

\begin{equation}
P_{a \rightarrow \gamma}= \biggl[\frac{B\,g_{a \gamma \gamma}}{2}
\biggr]^2 \,\,\frac{1}{q^2 +\Gamma^2 /4}\,\,\biggr[1 + e^{-\Gamma
L}-2 e^{-\Gamma\,L/2}\cos{(qL)}\biggr]
\end{equation}

where $L$ is magnet length and $\Gamma$ is the absorption
coefficient, which is zero in vacuum. The effective photon mass is
given by

\begin{equation}
 m_{\gamma} \approx \sqrt{\frac{4 \pi a N_e}{m_e}}=28.9
 \sqrt{\frac{Z}{A}\,\,\rho}\,\,\,eV
\end{equation}

 and the coherence condition is

\begin{equation}
 qL < \pi \Rightarrow \sqrt{m_{\gamma}^2 - \frac{2 \pi E_a}{L} } <
 m_a <  \sqrt{m_{\gamma}^2 + \frac{2 \pi E_a}{L} }
\end{equation}

The above condition is restored only for a narrow mass range
around $m_{\gamma}$, which for helium--4 can be adjusted by
changing the gas pressure as follows:

\begin{equation}
m_{\gamma}\,\,\,(eV) \approx \sqrt{0.02 \,\, \frac{P(mbar)}{T(K)}}
\end{equation}

\begin{figure}[t]
  \vspace{7.0cm}
  \includegraphics{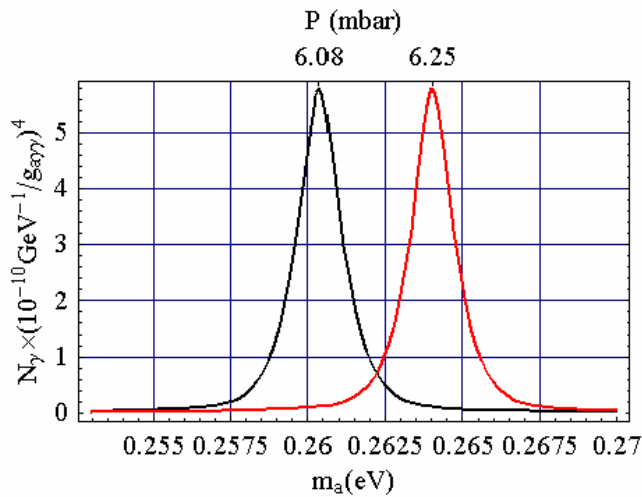}
  \caption{\it
    Probability of axion to photon conversion for two pressures.
    The step we used was about half this difference in order to
    scan fully the range of axion masses.
    \label{fig-6mbar} }
\end{figure}

As a matter of fact, every specific pressure allows to test a
specific axion mass. It is evident that, the higher the pressure,
the higher the photon effective mass, the higher the axion mass
under test. The transformation probability is shown in figure
\ref{fig-6mbar} for two pressures, namely 6.08 and 6.25 mbar,
indicating that the step has to be well below this difference
(0.17 mbar) in order to cover fully the axion mass range. In our
measurement program we used half this difference as step, namely
0.083 mbar. It is evident that for every step, there is a new
discovery potential.

Measurements with helium--4 have been already carried out up to a
pressure of 13.43 mbar, with small pressure steps. This is the
upper limit in pressure, before condensation effects take place.
This search tested the axion mass region up to 0.39 eV. The area
explored with helium--4 is designated in figure \ref{fig-ex-plot}.
Analysis is going on for these data and results will be published
in a forthcoming paper. CAST is currently upgraded in order to use
helium--3 as a buffer gas, allowing to increase the pressure up to
about 135 mbar and extending its sensitivity up to 1 eV axions.
The above CAST searches will allow to explore experimentally the
area of masses and coupling constants predicted by the axion
models, as it is shown in figure \ref{fig-ex-plot}. For higher
pressures there is a limitation coming not only from condensation
effects, but also from the photon absorption coefficient which
increases with pressure.

\section {Conclusions}

CAST searched for photons arising from axion conversion in a LHC
test magnet of 9.26 m length and 9 T magnetic field, for axion
masses less than 0.02 eV. Axions are expected to be produced in
the sun by the Primakoff process. The experiment obtained the best
experimental limit so far (an order of magnitude better than
previous experiments). Our result is for the first time better
than limits set by astrophysical arguments related to the
population of Horizontal Branch stars in globular clusters. This
population depends on their helium--burning lifetime. We have
searched for higher mass axions, up to 0.39 eV, by filling the
magnet bores with helium--4 buffer gas. Under these conditions,
photons acquire a small mass depending on pressure and coherence
condition between axions and photons is restored. Results from
this search will be published in a forthcoming paper. CAST is now
under upgrade, in order to fill the magnet bores with helium--3,
allowing us to explore axion masses up to 1 eV and testing the
range of $g_{a\gamma}$ - $m_a$ values anticipated from QCD axion
models and also the possible existence of large extra dimensions
\cite{extra-dims}. Axions of this range of masses could be
candidates for a hot dark matter component of the universe
\cite{hdm}.

\section {Acknowledgments} I would like to thank K.Zioutas,
A.Liolios, I.Savvidis and T.Papaevangelou for fruitful
discussions. I would like also to thank E. Ferrer Ribas and
M.Kuster for suggestions/comments. We acknowledge support from
NSERC (Canada), MSES (Croatia), CEA (France), BMBF (Germany) under
the grant numbers 05 CC2EEA/9 and 05 CC1RD1/0, the Virtuelles
Institut f\"ur Dunkle Materie und Neutrinos -- VIDMAN (Germany),
GSRT (Greece), RFFR (Russia), CICyT (Spain), NSF (USA) and the
helpful discussions within the network on direct dark matter
detection of the ILIAS integrating activity (Contract number:
RII3-CT-2003-506222). We thank CERN for hosting the experiment and
for the contributions of CERN staff.

\end{document}